\documentclass{evn2004}
\usepackage{txfonts}
\begin{document}
   \title{How to improve the High Frequency capabilities of SRT}

   \author{T. Pisanu\inst{1}, M. Morsiani\inst{1}, C. Pernechele\inst{2}, F. Buffa\inst{2}
          \and
          G. Vargiu\inst{1}
          }

   \institute{Institue of Radio Astronomy (CNR), via P. Gobetti 101, 40129 BO, Italy
         \and
             Astronomical Observatory of Cagliari, loc. Poggio dei Pini strada 54, 09012 Capoterra(CA), Italy
             }

   \abstract{
The SRT (Sardinia Radio Telescope) is a general purpose, fully
steerable, active surface equipped, 64 meters antenna, which is in
an advanced construction state near Cagliari (Sardinia - Italy).
It will be an antenna which could improve a lot the performances
of the EVN network, particularly at frequencies higher than 22
GHz.

The main antenna geometry consist of a shaped reflector system
pair, based on the classical parabola-ellipse Gregorian
configuration. It is designed to be able to operate with a good
efficiency in a frequency range from 300 MHz up to 100 GHz. This
frequency range, is divided in two parts which define also two
antenna operational modes, one up to 22 GHz with a minimal amount
of accessory instrumentation, and the other up to 100 GHz with a
full complement of instrumentation.

The goal is to make it possible to build a telescope operable up
to 22 GHz, and then upgrade it at a future date to operate at
frequencies up to 100 GHz.

In order to get these goals, the SRT Metrology group is studying
and developing different types of strategies, instrumentation, and
techniques for measuring and reducing the various components of
pointing and efficiency errors, taking advantage also from
experiences developed in other radio telescopes, like GBT (Green
Bank Telescope, USA), LMT (Large Millimiter Telescope, MEX), and
IRAM (Institut de Radio Astronomie Millimetrique, Fr).

Many of those system will be installed and tested at the 32 meters
radio-telescope in Medicina (Bologna), before of their
implementation on SRT.

   }

   \maketitle
%

\section{Introduction}

The causes which mainly contribute to a degradation in pointing
and gain efficiency of a Radio Telescope, can be divided in two
main classes: repeatable errors (for example mechanical errors
during initial alignment, gravity deformations) and more critical,
non-repeatable errors (thermal gradients and wind effects).

The performances of the antenna in pointing and surface accuracy,
clearly depends on environmental conditions, better with benign
conditions, and degraded when the environment is severe.

SRT weights more than 3000 ton and has a diameter of 64 meters.
According to von Hoerner (\cite{horner}) the maximum rms surface
value for a millimeter-wavelength telescope of 64 meters, is $\sim
2 mm$, which means a $\lambda_{min}$ of $\sim 9 GHz$ using the
empirical rule of $\lambda_{min}\sim 16 \sigma$.

To have a good efficiency at 100 GHz, the root-mean-square
accuracy of the reflector surfaces, should be of $\epsilon \sim
\lambda/20 \sim 190 \mu m$, and the precision for non-repeatable
pointing at 100 GHz, should be a tenth of the beam width, i.e., 1
arcsec in good environmental conditions, which means without solar
radiation, no precipitation, air temperature between $-10^{o}$ C
and $40 ^{o} C$, a temperature drift $< 10^{o} C/h$ and an
humidity $< 90 \%$.

To compensate those errors, we are testing different sub-systems
for measuring and compensating the causes.

The sub-systems that we are studying are: a \textbf{laser with a
PSD} (Position Sensing Device) system for measuring the
sub-reflector position, a \textbf{temperature probes} network for
measuring and predicting thermal deformations on the supporting
structure, a \textbf{star tracker} system for assisting pointing
model upgrade and source tracking, \textbf{anemometers} and
pressure probes for measuring wind effects, a \textbf{FEM model}
of the antenna for assisting the metrology development and for
predicting wind and thermal deformations, \textbf{two
inclinometers} for measuring aliadade and rail track deformations
and a \textbf{laser ranging} system for measuring structural
dimensions.


\section{Sub-reflector alignment}


To achieve the required pointing accuracy for the SRT 64-meter telescope,
the position of the subreflector must be accurately known relative to
the vertex of the main reflecting surface.

A laser metrology system with five degrees of freedom is the ideal
technical solution.

The system shall consist of a stationary laser and a 20 meters far
away head sensor, able to measure its more important degrees of
freedom.

A system which could satisfy our requirements, is that produced by Apisensor (www.apisensor.com),
the API 6D Laser Measuring System, with a linear range up to 25 meters, a linear accuracy of 1 p.p.m,
a straightness accuracy of $1 \mu m$ and an angular accuracy of 1.0 arc-second.

We are also testing a less expensive system, using a solid state
laser, with a PSD from Duma Optronics (www.duma.co.il), the model
AlignMeter PC. In this case, the head sensor has two $10 \times 10
mm$ dual axis silicon Lateral Effect PSD, one for measuring the
translations perpendicular to the laser beam direction and the
other one, which is in the focal plane of a lens, measure the
angular deviations.

The position resolution is better than $\pm 1\mu m$, and the angle
resolution is $~ \pm 2 arcsec$, the operational spectral range
goes from 300 to 1100 nm.


\section{Temperature monitoring}

A network of temperature probes distributed on the whole
structure, (alidade, backstructure of the primary mirror, primary
panels, quadrupod and secondary mirror) has been also foreseen.

Because we have abandoned the originally planned thermal
insulation and forced ventilation of the reflector surface, as is
at the 30 m IRAM telescope(\cite{Baars}), this network of
temperature sensors is very crucial.

The optimal position of each probe will be set through a FEM
analysis software which, according to the measured temperatures,
will drive the active surface actuators in a open loop fashion.

The sensors which we acquired and tested in the laboratory, are
Negative Thermistors NTC from YSI, model GEM 55036 with a
Resistance $\Omega$ at $25^{o} C$ of $10 k\Omega$ and a tolerance
interchangeability  of $\pm 0.1^{o} C$ from 0 to $70^{o} C$.

We performed few laboratory test, introducing our sensors in a
cooler and measuring their responses from -20 up to $70^{o} C$.
The results are in a very good accordance with predictions.

\section{Optical Star Tracker}

The pointing model and the tracking, will be supported by an
optical star tracker (OST), which we are planning to develop and
to use also in daylight situations.

In order to have enough sky coverage it will be realized by using
a commercial Maksutov-Cassegrain optical telescope with a primary
diameter of 180 mm and an aperture of f/10. The CCD will be a
peltier cooled 512x512 pixels with 20 micron of pixel size. The
covered field of view results of 19 x 19 arcminutes, while the
plate scale is of 2.3 arcsec/px. The latter seems to be well
compatible with the typical atmospheric conditions of the site.

Initially we are planning to use the OST only for monitoring
purposes, of the pointing model and of the antenna servo system.
The last step will be one of realize the auto-guide during day
time, using a IR filter able to cut off the visible wavelenghts.

\section{Pressure probes}

The wind pressure sensor system hardware is required for good
pointing accuracy whenever the wind is in excess of "precision"
wind conditions.

The accuracy of this hardware will greatly impact the pointing,
focus, and surface accuracies of the telescope.

The wind pressure sensor hardware shall operate at full accuracy
when subjected to "Normal" operating conditions, i.e., an ambient
temperature range of -10  C to +40 C, with humidity   90\%. When
operating in "Extreme" conditions, i.e., -15 to +50 $^{o}C$,
degraded accuracy is permitted. All components shall survive
(non-operating) in temperatures down to  30 $^{o}C$.

From the FEM model, we have found that for having a good are
necessary a total of 152 sensors are used. This number of sensors
gives significant redundancy and hence high availability to system

\section{The Finite Element Model}

The improvement in the predictions from FEM software simulations,
involved us to use it as a diagnostic tool for knowing the more
critical points in the antenna structure and for knowing the
thermal and wind effects on it.

There are many studies related to the use of FEM simulations for
predicting and correcting non-systematic errors in
radio-telescopes.

We already have at home the FEM model of the antenna, developed
for us by BCV consulting with the ANSYS software, for evaluating
the compatibility between the preliminary project and the
executive project.

Once that the structural critical point have been selected, we can
use the FEM together with the measured temperature and wind
pressures, for predicting and correcting the pointing and
efficiency errors.

\section{Inclinometers}

The tilt meter subsystem will consist of two bi-axial tilt meters,
with one tilt meter located near each of the elevation bearings.

The purpose of the tilt meters is to measure changes (due to wind
and temperature) in the orientation of the elevation axle relative
to the encoders. Once these rotations are measured, a portion of
the pedestal's pointing error can be calculated and then removed.

A described before, the metrology system, uses wind pressure
sensors, thermal sensors, and subreflector laser metrology to
compute and correct for pointing errors.

The information from the tilt meters will be used mainly as a
"sanity check" and redundant back-up system.

\section{Laser RangeFinders}

Finally, the possibility to install laser rangefinders for
monitoring the primary reflector surface deformation, and if
necessary the pointing with the help of triangulation algorithm,
as foreseen also for the GBT, is under study.

%
\section{Conclusions}

The possibility to use SRT up to 100 GHz, will depend from the
performances of the metrology system mounted on it.

We are involved into the study of different sub-systems,  for
measuring and correcting the structural deformations produced by
gravity, temperature gradients and wind.

Few of those sub-systems are redundant, because of their unknown
performances and so we are experimenting more than one in such a
way that we could use the more performing.

\end{document}